# ForecastTB—An R Package as a Test-Bench for Time Series Forecasting—Application of Wind Speed and Solar Radiation Modeling

**Neeraj Dhanraj Bokde [1,\*], Zaher Mundher Yaseen [2]** 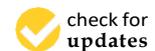 **and Gorm Bruun Andersen [1]**

[1] Department of Engineering—Renewable Energy and Thermodynamics, Aarhus University, 8000 Aarhus, Denmark; gba@eng.au.dk

[2] Sustainable Developments in Civil Engineering Research Group, Faculty of Civil Engineering, Ton Duc Thang University, Ho Chi Minh City, Vietnam; yaseen@tdtu.edu.vn

\* Correspondence: neerajdhanraj@eng.au.dk; Tel.: +91-902-841-5974

**Abstract:** This paper introduces an R package ForecastTB that can be used to compare the accuracy of different forecasting methods as related to the characteristics of a time series dataset. The ForecastTB is a plug-and-play structured module, and several forecasting methods can be included with simple instructions. The proposed test-bench is not limited to the default forecasting and error metric functions, and users are able to append, remove, or choose the desired methods as per requirements. Besides, several plotting functions and statistical performance metrics are provided to visualize the comparative performance and accuracy of different forecasting methods. Furthermore, this paper presents real application examples with natural time series datasets (i.e., wind speed and solar radiation) to exhibit the features of the ForecastTB package to evaluate forecasting comparison analysis as affected by the characteristics of a dataset. Modeling results indicated the applicability and robustness of the proposed R package ForecastTB for time series forecasting.

## 1. Introduction

Decision making is one of the most crucial tasks in many domains and often decisions are based on the most accurate forecast available in the respective domains. A large number of areas, such as energy [1,2], economics [3], infrastructure [4,5], health [6,7], agriculture [8,9], defense [10], education [11,12], technology [13,14], geo-science [15], climate [16] and structural engineering [17] among several others, are looking forward to benefits that can be achieved with time series forecasting. Time series are consecutive sequences of values ordered with respect to time. Statistically, it is represented by the theory of stochastic processes [18]. The notable feature of time series data is the time-based correlation between values such that the probability of occurrence of a value depends on future or past observations [19,20].

Standardizing the process of the forecasting methods comparison is associated with numerous challenges [21]. The apparent concern is the nature of the time series and selection of appropriate methods such as the complete variation from seasonality, trends, and random parameters, that should be handled with each method. For instance, one forecasting model may perform better for a seasonal time series but show poor accuracy as compared to other models for a trendy or a random one [22]. Besides, interpretations might be affected by the error metric selection, for example, Root mean square error (RMSE), as different metrics have different objectives [23].

The ForecastTB package can be helpful for professionals and researchers working in the field of data science and forecasting analysis. The salient features of the ForecastTB package are as follows:

- **Reduction in efforts and time consumption:** The ForecastTB package is designed to reduce the efforts and time consumption for the time series forecasting analysis. It avoids the repetitive steps in the analysis and leads to the promising comparative results report generation.
- **Truthful comparison assurance:** The ForecastTB package ensures a truthful and unbiased comparison of forecasting methods. Hence, this package may be considered a reliable tool for forecasting models based on industrial reports generation or scientific publications.
- **Reproducible research:** Along with unbiased comparisons, the ForecastTB package provides ease in reproducible research with minimum efforts. In other words, the forecasting comparison can be reproduced several times easily with the help of the ForecastTB package.
- **Stepping stone in machine learning automation:** Forecasting methods play a very important role in machine learning applications [24]. The ForecastTB package aims to evaluate the best performing forecasting method for a given time series dataset and this can be presented as a stepping stone in machine learning automation modeling. For example, on changing nature and patterns of the time series dataset, a machine learning application could automatically replace the existing forecasting methods based on the output of the ForecastTB package.
- **A handy tool:** The ForecastTB package is a handy tool, especially for researchers who are not comfortable with computer coding, since it is a plug-and-play module based package. A very simple syntax leads to very impressive and accurate forecasting comparison analysis.

This paper demonstrates the inspection of the ForecastTB package as a testbench for the comparative study of different forecasting methods [25]. The motivation behind this package is another R package, named imputeTestbench [26,27], which demonstrated great success in performing comparison analysis with time series imputation methods in several research studies [28–31]. The ForecastTB package aims at providing an evaluation toolset that overcomes the challenges of discovering the most suitable forecasting method along with a detailed comparative analysis. This package allows simulating random possibilities of different strategies including Monte-Carlo simulation. Values forecasted with several different methods are then compared with a user defined error metric. A couple of plotting functions are provided for overall forecast evaluation visualization. Besides, demonstration examples and case studies on natural time series datasets are discussed to showcase the usage of ForecastTB package.

## 2. Overview of ForecastTB

The ForecastTB package is a plug-and-play structured module as shown in Figure 1. It is used to compare the forecasting methods, which begins by forecasting time series with distinct strategies. Then the prediction accuracies are evaluated with the error metrics for all methods for several repetitions. The prediction_errors() function is employed for a forecasting method comparative evaluation with the consideration of various input parameters. It returns an object, which is the basic module in the package. Further this module can be updated with new methods and other parameters with append_() and choose_() functions. The Monte_Carlo() function is a further extension of the prediction_errors() module to compare distinct methods for randomly selected patches of the input time series. The remaining two functions, plotMethods() and plot_circle(), are used to visualize forecasted results and error summaries for the chosen methods. All functions, as a set of modules, are based on and connected with the object provided by the prediction_errors() function. Hence, the framework of the ForecastTB package is version-control-friendly. It means, in the future, new features in the next versions of the package, can be easily introduced.

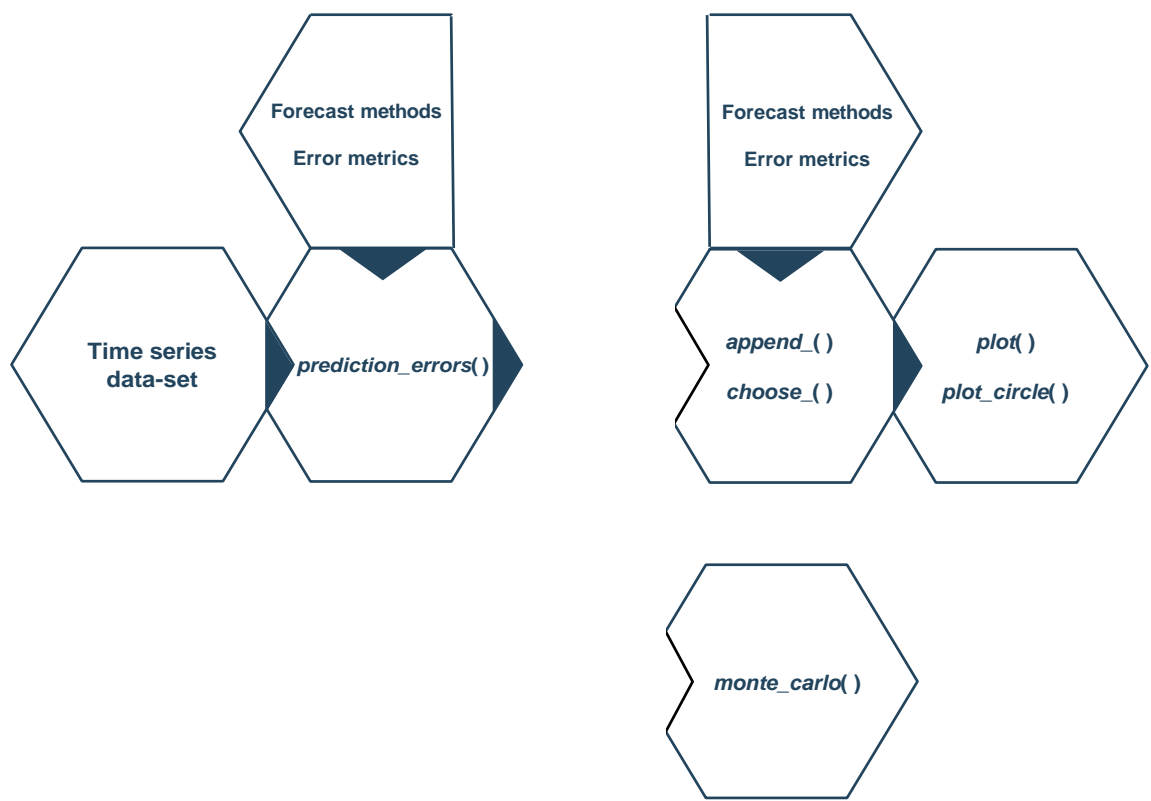

**Figure 1.** The plug-and-play module of the ForecastTB package.

Dependencies include additional packages for data manipulation (reshape2, [32]; stats, [33]), graphing (ggplot2, [34]; circlize, [35]; graphics, [33]; gridExtra, [36]; RColorBrewer, [37]), and forecast (forecast, [38]; PSF, [39–41]; decomposedPSF, [42]; methods, [33]).

Recently, a package in R was proposed for streamlining time series forecasting with limited facilities and features, named as predtoolsTS [43,44]. This tool assists in forecasting with the automated Auto-Regressive Integrated Moving Average (ARIMA) model [38] and only regression type algorithms from the caret [45] package in the closed environment. Beyond this, there is no facility to append new forecasting methods and updating the comparison environment as per the user's need. On the contrary, the proposed ForecastTB package provides such multiple facilities along with the possibilities of introducing new features with a simple plug-and-play structure as discussed in the article.

## 2.1. The prediction_errors() function

The prediction_errors() is a primary function that compares the performance of different forecasting methods depends on various parameters supplied by user. The default methods included in prediction_errors() are ARIMA from the forecast package [38] and the Pattern sequence based forecast (PSF) method from the PSF package [39,40]. This statistical method, ARIMA, is most widely and frequently used in time series forecasting. It is comfortably modeled as compared to more complex approaches. Moreover, to avoid further complexity, the auto.arima() function in the forecast package [38] and psf() from the PSF package are used, which automatically estimates the model parameters required for the given dataset. With the following examples, we show the possibility of adding extra and user-defined methods as needed.

Following are the arguments of the prediction_errors() function:

```
a <- prediction_errors(data = data, nval = 12,
ePara = c('RMSE', 'MAE'), ePara_name = c('RMSE', 'MAE'),
Method = c('ARIMA', 'PSF'), MethodName = c('ARIMA', 'PSF'),
```

strats = 'Recursive', dval = length(data),
append_ = 0)

**data:**

A ts (stats) as input dataset (numeric object) to be evaluated. It is a time varied dataset for comparison of performance and accuracy of forecasting methods.

**nval:**

An integer value to decide a number of values to be predicted while comparing the forecasting methods. The default value is 12.

**ePara:**

The forecasted and observed values are compared with error metrics. The root-mean-squared error (RMSE), mean absolute error (MAE), and mean absolute percent error (MAPE) [46] are the default error metrics included in ForecastTB, but the individual metric can be opted using ePara = 'rmse', 'mae', or 'mape'. The error metrics are defined as:

$$RMSE = \sqrt{\frac{1}{N} \sum_{i=1}^{N} |X_i - \hat{X}_i|^2} \tag{1}$$

$$MAE = \frac{1}{N} \sum_{i=1}^{N} |X_i - \hat{X}_i| \tag{2}$$

$$MAPE = \frac{1}{N} \sum_{i=1}^{N} \frac{|X_i - \hat{X}_i|}{X_i} \times 100\%, \tag{3}$$

where $X_i$ and $\hat{X}_i$ are the observed and forecasted data at time t, respectively. N is the number of data for forecast evaluation.

Additionally, a new error metric of the user's choice can be introduced in the following manner: ePara = c("errorparametername"), where errorparametername is an external function which returns desired error values for original and forecasted values. It is discussed in detail in the following sections.

**ePara_name:**

A string of characters for the names of error metrics provided in ePara. In the prediction_errors() function, RMSE, MAE and MAPE are the default names for the error parameters.

**Method:**

A string with method names that are used to forecast values with ARIMA, PSF and user-defined methods. The ARIMA is a default method until the user changes it. For instance, Method = c("PSF") will use only the PSF with the prediction_errors() function. New forecasting methods can be included in prediction_errors() function with Method = c("test1(data, nval)") vector, where test1 is an external function which forecasts nval number of future values for input data and returns the forecasted values. Such multiple functions can be added in the Method parameter. The external function is expected to provide forecast methods using the recursive strategy.

**MethodName:**

A string of characters provided by user as one or more forecast methods passed to Method. The default list of name of method in the prediction_errors() function is ARIMA.

**strats:**

A character string of names of forecasting strategies used for time series forecasting. The strategies used in the current version of the prediction_errors() function are Recursive and DirRec. In the recursive approach, the first value is predicted with a prediction model and then appended to the data-set. Fits to the same model are then used to predict the next value. In the DirRec approach, the first value is predicted with a model and appended it to the data-set. A new model is then fitted and used to predict the next value and so on. The default strategy in the prediction_errors() function is the Recursive one.

**append_:**

A flag that suggests if the function is used to append to another instance. The default flag is 1, which indicates newly added methods are compared with the default methods. When the flag is 0, all methods, excluding default methods, will be compared.

**dval:**

An integer that specifies the length of the data subset to be used for forecasting. The default value is the length of the data parameter.

With the default entities for all arguments, the prediction_errors() function generates an S4 object of class prediction_errors as the error profile for the forecasting methods. This prediction_errors() function returns two slots as output. The first slot is output, which provides Error_Parameters, indicating error values for the forecasting methods and error parameters defined in the framework, and Predicted_Values as values forecasted with the same foreasting methods. Further, the second slot is parameters, which returns the parameters used or provided to prediction_errors() function. This is illustrated in the following example:

```
#'AirPassengers' is a sample dataset in CRAN
prediction_errors(data = AirPassengers)
```

```
## An object of  class  "prediction_errors"
## Slot "output":
## $Error_Parameters
##                RMSE        MAE        MAPE exec_time
## ARIMA  53.357458  45.662500   9.647511    1.248623
##
## $Predicted_Values
##                    1         2         3         4         5         6         7
## Test values  417.0000  391.0000  419.0000  461.0000  472.0000  535.0000  622.0000
## ARIMA        425.5702  432.2525  489.0878  499.4493  539.9511  539.7919  536.0011
##                    8         9        10        11        12
## Test values  606.000  508.0000  461.000  390.0000  432.0000
## ARIMA        525.408  492.9301  486.233  461.6871  470.2665
##
##
## Slot "parameters":
## $data
##       Jan Feb Mar Apr May Jun Jul Aug Sep Oct Nov Dec
## 1949  112 118 132 129 121 135 148 148 136 119 104 118
## 1950  115 126 141 135 125 149 170 170 158 133 114 140
## 1951  145 150 178 163 172 178 199 199 184 162 146 166
## 1952  171 180 193 181 183 218 230 242 209 191 172 194
```

```
## 1953 196 196 236    235 229 243 264 272 237 211 180 201
## 1954 204 188 235    227 234 264 302 293 259 229 203 229
## 1955 242 233 267    269 270 315 364 347 312 274 237 278
## 1956 284 277 317    313 318 374 413 405 355 306 271 306
## 1957 315 301 356    348 355 422 465 467 404 347 305 336
## 1958 340 318 362    348 363 435 491 505 404 359 310 337
## 1959 360 342 406    396 420 472 548 559 463 407 362 405
## 1960 417 391 419    461 472 535 622 606 508 461 390 432
##
## $nval
## [1] 12
##
## $ePara
## [1] "RMSE" "MAE"    "MAPE"
##
## $ePara_name
## [1] "RMSE" "MAE"    "MAPE"
##
## $Method
## [1] "ARIMA"
##
## $MethodName
## [1] "ARIMA"
##
## $Strategy
## [1] "Recursive"
##
## $dval
## [1] 144
```

## 2.2. *The append_() and choose_() functions*

The append_() and choose_() are functions to add new forecasting methods and choose the selected methods to and within the prediction_errors object. These functions make ForecastTB a handy plug-and-play tool for forecasting analysis. The append_() function has the following arguments:

```
append_(object = a,
Method = c("test2(data,nval)"), MethodName = c('Test2'),
ePara = c('RMSE', 'MAE'), ePara_name = c('RMSE', 'MAE'))
```

**object:**

A prediction_errors object with two slots, output and parameters returned by the prediction_errors() function.

Method, MethodName, ePara and~ePara_name:

These parameters are with similar meaning as that supplied to the prediction_errors() function. The ePara and ePara_name parameters of the append_() function automatically get synced with the prediction_errors() function. The append_() function returns the error profile for the forecasting methods along with newly added ones in the form of an updated prediction_errors object.

Further, the choose_() function helps the user to manipulate the prediction_errors object, by allowing selection of the desired methods existing in the object. The choose_() function has the following argument:

choose_(object)

object:

A prediction_errors object returned by the prediction_errors() or append_() function. The syntax of the choose_() function is discussed in the following example. The variable c is an object with three forecasting methods comparisons. After provide the c object to the choose_() function, it returns the following output showing names of forecasting methods attached in the c object and asks the index number of method to be removed from the same object. Finally, it stores the newly updated object in the variable d.

```
## > d <- choose_(object = c)
## Following are the~methods attached with the~object:
##            [,1]     [,2]    [,3]
## Indices  "1"      "2"     "3"
## Methods "ARIMA" "DPSF" "xyz"
## Enter the~indices of methods to remove:2
```

After removing on of the methods with the choose_() function, a new object d is returned as follows:

```
## > d@output
## $Error_Parameters
##              RMSE       MAE       MAPE exec_time
## ARIMA 2.3400915 1.9329816 4.2156087 0.1328349
## xyz    4.0909453 3.3250000 7.1517742 0.4336619
##
## $Predicted_Values
##                    1        2        3        4        5        6        7        8
## Test values 39.40000 40.90000 42.40000 47.80000 52.40000 58.00000 60.70000 61.80000
## ARIMA        37.41933 37.69716 41.18252 46.29926 52.24804 57.10696 59.71674 59.41173
## xyz          44.20000 39.80000 45.10000 47.00000 54.10000 58.70000 66.30000 59.90000
##
##                    9       10       11       12
## Test values 58.20000 46.7000 46.60000 37.80000
## ARIMA        56.38197 51.4756 46.04203 41.52592
## xyz          57.00000 54.2000 39.70000 42.80000
```

In this way, the append_() and choose_() functions provide features to manage the desired forecasting methods in the prediction_errors object with simple and single lined syntax. Besides, these functions avoid repetitive computation and execution of the methods which are already attached in the prediction_errors object.

## 2.3. The plot.prediction_errors() and plot_circle() functions

The plot.prediction_errors() and plot_circle() functions allow users to plot the S4 object of the prediction_errors class. The plot.prediction_errors() is an S3 method, in which, along with the prediction_errors object, it optionally accepts other variables in used to describe plots in R, such as, legends, axes, labels, and so forth. This function plots a bar-chart of error values and execution time and line plots of forecasted values with significant color changes for different forecasted methods. The plot generated plot.prediction_errors() consists of a dynamic margin size with which forecasted values can be included if required. The **plot.prediction_errors()** function syntax and usage are discussed below:

**plot.prediction_errors(object = a, ...)**

Furthermore, the plot_circle() function provides a polar plot for showing forecasted values, especially if these are seasonal values. This plot shows how forecasted observations are behaving on an increasing number of seasonal time horizons. The syntax of the plot_circle() function is similar to that of plot.prediction_errors() one, as shown below:

**plot_circle(object = a, ...)**

### 2.4. The monte_carlo() function

Monte-Carlo sampling is a very popular strategy to compare the performance of forecasting methods. In this strategy, multiple patches of a dataset are selected randomly and then tests of the performance of the forecasting methods are executed. It returns the average of error values obtained for each data patches. The most notable feature of the Monte-Carlo strategy is that it ensures an accurate comparison of forecasting methods and avoids the biased results which might be obtained by chance.

The monte_carlo() function has the following arguments:

**monte_carlo(object, size, iteration, fval, figs)**

**object:**

An S4 object of the prediction_errors class with two slots, output and parameters returned by the prediction_errors(), append_() or choose_() functions. The monte_carlo() function tunes itself based on parameters returned by the object.

**size:**

An integer value representing the length of the patches from the time series dataset used in the Monte Carlo strategy. The input dataset is referred from the prediction_errors object.

**iteration:**

A numeric value representing the iterations to be used by the Monte-Carlo strategy.

**fval:**

A flag, when it is on (fval = 1), the monte_carlo() function return the output along with the forecasted values for each iteration. The default value of fval is 0, that returns only the error values in RMSE in each iteration.

**figs:**

A flag to plot with the plot.prediction_errors() function for each iteration in Monte-Carlo strategy when figs = 1. The default value of figs is 0 and it does not return any plots.

This way, the monte_carlo() function returns the error values along with the mean values for the provided forecasted methods in each iteration of the Monte-Carlo strategy. This function can be used as a quick tool to create an intuition of best-suited forecasting method for a given time series dataset.

## 3. A Case Study: Performance of Forecasting Methods on Standard Natural Time Series

This section presents a case study to introduce the use of the ForecastTB package as a test-bench to compare the performance of time series forecasting methods. The methods used in the case study are the auto.arima(), ets(), psf(), and lpsf() that are frequently used R functions for time series forecasting. The two standard datasets used in this example are nottem and sunspots. The nottem is the average air temperatures recorded at Nottingham Castle in degrees Fahrenheit from 1920 to 1930 (datasets package, [33]) Similarly, the sunspots is mean relative sunspot numbers from 1749 to 1983 (datasets package, [33]). Both datasets are measured on monthly basis.

In the following examples, the whole comparison analysis and visualization are done for the nottem dataset, but only the error values are shown for the sunspots dataset. The following are the two functions that represent forecasting methods. These methods include LPSF and PSF methods from the decomposedPSF [42] and PSF [39] packages. These functions consume time series dataset and nval as input parameters and return the forecasted values as a string.

```r
library(decomposedPSF)
test1 <- function(data, nval){
return(lpsf(data = data, n.ahead = nval))
}

library(PSF)
test2 <- function(data, nval){
    a <- psf(data = data, cycle = 12)
    b <- predict(object = a, n.ahead = nval)
return(b)
}
```

The following code chunk is the main function in the ForecastTB package, in which performance of ARIMA along with the above mentioned (PSF and LPSF) methods are compared for the nottem time series for next 12 months forecast. The output of the function is saved in the variable a1 as an S4 object of prediction_errors class.

```r
a1 <- prediction_errors(data = nottem, nval = 12,
Method = c("test1(data, nval)", "test2(data, nval)"),
MethodName = c("LPSF","PSF"),
append_ = 1)
a1
```

```
## An object of class "prediction_errors"
## Slot "output":
## $Error_Parameters
##                 RMSE       MAE      MAPE exec_time
## ARIMA 2.3400915 1.9329816 4.2156087 0.2290468
##  LPSF  5.3525306 4.5916667 9.6590830 0.2280521
##  PSF   2.2454324 1.9450000 4.1462600 0.1007819
##
## $Predicted_Values
##                   1         2         3         4         5         6         7
## Test values 39.40000  40.90000  42.40000  47.80000  52.40000  58.00000  60.70000
## ARIMA       37.41933  37.69716  41.18252  46.29926  52.24804  57.10696  59.71674
## LPSF        38.55000  37.10000  34.90000  43.15000  45.50000  52.40000  55.95000
## PSF         38.38000  37.56000  41.80000  45.46000  52.96000  57.14000  61.68000
```

```
##                          8         9        10        11        12
## Test values 61.80000  58.20000  46.7000  46.60000  37.80000
## ARIMA        59.41173  56.38197  51.4756  46.04203  41.52592
## LPSF         61.45000  60.40000  57.2500  49.75000  42.60000
## PSF          59.20000  56.68000  49.6000  42.56000  40.38000
##
##
## Slot "parameters":
## $data
##         Jan  Feb  Mar  Apr  May  Jun  Jul  Aug  Sep  Oct  Nov  Dec
## 1920  40.6 40.8 44.4 46.7 54.1 58.5 57.7 56.4 54.3 50.5 42.9 39.8
## 1921  44.2 39.8 45.1 47.0 54.1 58.7 66.3 59.9 57.0 54.2 39.7 42.8
## 1922  37.5 38.7 39.5 42.1 55.7 57.8 56.8 54.3 54.3 47.1 41.8 41.7
## 1923  41.8 40.1 42.9 45.8 49.2 52.7 64.2 59.6 54.4 49.2 36.3 37.6
## 1924  39.3 37.5 38.3 45.5 53.2 57.7 60.8 58.2 56.4 49.8 44.4 43.6
## 1925  40.0 40.5 40.8 45.1 53.8 59.4 63.5 61.0 53.0 50.0 38.1 36.3
## 1926  39.2 43.4 43.4 48.9 50.6 56.8 62.5 62.0 57.5 46.7 41.6 39.8
## 1927  39.4 38.5 45.3 47.1 51.7 55.0 60.4 60.5 54.7 50.3 42.3 35.2
## 1928  40.8 41.1 42.8 47.3 50.9 56.4 62.2 60.5 55.4 50.2 43.0 37.3
## 1929  34.8 31.3 41.0 43.9 53.1 56.9 62.5 60.3 59.8 49.2 42.9 41.9
## 1930  41.6 37.1 41.2 46.9 51.2 60.4 61.6 57.0 50.9 43.0 38.8
## 1931  37.1 38.4 38.4 46.5 53.5 58.4 60.6 58.2 53.8 46.6 45.5 40.6
## 1932  42.4 38.4 40.3 44.6 50.9 57.0 62.1 63.5 56.3 47.3 43.6 41.8
## 1933  36.2 39.3 44.5 48.7 54.2 60.8 65.5 64.9 60.1 50.2 42.1 35.8
## 1934  39.4 38.2 40.4 46.9 53.4 59.6 66.5 60.4 59.2 51.2 42.8 45.8
## 1935  40.0 42.6 43.5 47.1 50.0 60.5 64.6 64.0 56.8 48.6 44.2 36.4
## 1936  37.3 35.0 44.0 43.9 52.7 58.6 60.0 61.1 58.1 49.6 41.6 41.3
## 1937  40.8 41.0 38.4 47.4 54.1 58.6 61.4 61.8 56.3 50.9 41.4 37.1
## 1938  42.1 41.2 47.3 46.6 52.4 59.0 59.6 60.4 57.0 50.7 47.8 39.2
## 1939  39.4 40.9 42.4 47.8 52.4 58.0 60.7 61.8 58.2 46.7 46.6 37.8
##
## $nval
## [1] 12
##
## $ePara
## [1] "RMSE" "MAE"  "MAPE"
##
## $ePara_name
## [1] "RMSE" "MAE"  "MAPE"
##
## $Method
## [1] "ARIMA"              "test1(data, nval)" "test2(data, nval)"
##
## $MethodName
## [1] "ARIMA" "LPSF"  "PSF"
##
## $Strategy
## [1] "Recursive"
##
## $dval
```

## [1] 240

The object a1 shows the comparison of ARIMA, PSF, and LPSF methods in addition to the parameters provided by users and the defaults ones, whenever not provided. Imagining a case that the user wishes to attach a new method in the object a1. Though this could be done with the prediction_errors() function, as discussed above, this will lead to the execution of all methods which were already available in the object. To avoid such redundant calculations and executions, the append_() function could be used. The Error, Trend, Seasonal (ETS) method from the forecast [38] package is framed in a function as follows:

```r
library(forecast)
test3 <- function(data, nval){
  b <- as.numeric(forecast(ets(data), h = nval)$mean)
return(b)
}
```

The object a1 is updated with ETS method into object c1 with the append_() function and returns the error parameters as shown below:

```r
c1 <- append_(object = a1, Method = c("test3(data,nval)"), MethodName = c('ETS'))
c1@output$Error_Parameters
```

```
##                  RMSE          MAE         MAPE    exec_time
##   ARIMA     2.3400915    1.9329816    4.2156087    0.2290468
##   LPSF      5.3525306    4.5916667    9.6590830    0.2280521
##   PSF       2.2454324    1.9450000    4.1462600    0.1007819
##   ETS      31.41914256  28.87699844  56.31295348  0.04888916
```

Further, the object c1 is plotted with the S3 method plot.prediction_errors() as follows and are shown in Figure 2. This figure shows the forecasting errors (bar plot) and forecasted values (line plot) for all methods attached in object c1.

```r
d1 <- plot(c1)
```

From Figure 2, it can be observed that the performance of the ETS method is not satisfactory for the nottem time series dataset. In this case, the user can omit the unwanted method from object d1 with the choose_() function as shown below. After providing the object as input to the choose_() function, it returns the names of forecasting methods available in the prediction_errors object with respective indices, and asks the user to enter the indices of the methods to be removed from the object. In the following code chunk, index 4 for the ETS method is provided by the user and this results in a new object e1, which is nothing but the object d1 without the ETS method. The choose_() function permits the user to remove multiple numbers of methods from the object, where the user needs to enter multiple numbers in a vector format.

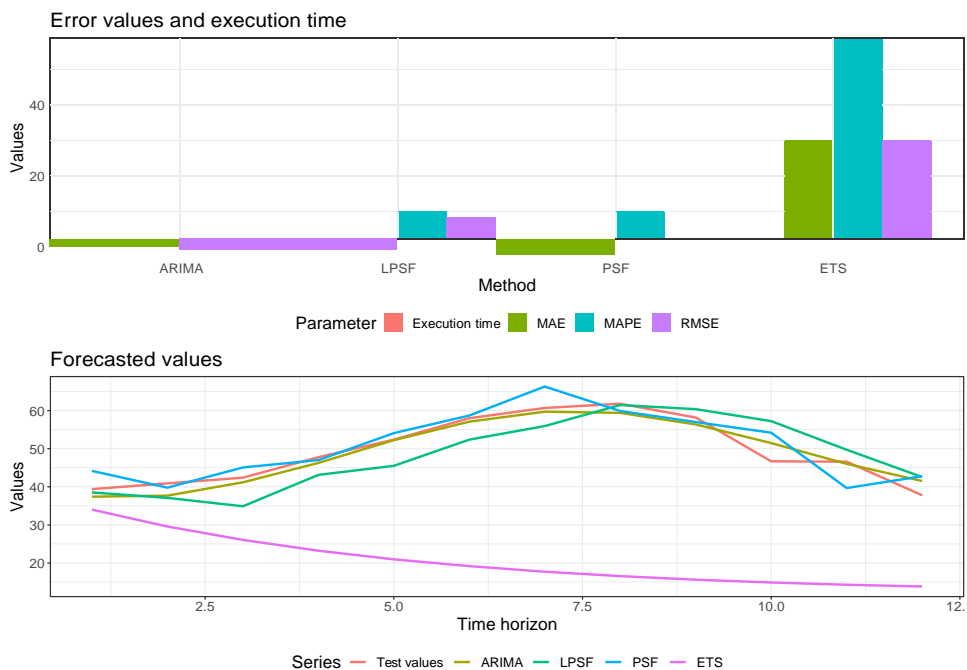

**Figure 2.** Forecasting errors (in bar plot) and forecasted values (in line plot) for methods attached in the object c1.

```
# > e1 <- choose_(object = c1)
# Following are the~methods attached with the~object:
#             [,1]     [,2]     [,3]   [,4]
# Indices    "1"      "2"      "3"    "4"
# Methods "ARIMA"  "LPSF"   "PSF"  "ETS"
#
# Enter the~indices of methods to remove:4
#
# > e1@output$Error_Parameters
#              RMSE          MAE exec_time
# ARIMA 2.5233156 2.1280641 0.1963789
# LPSF  2.3915796 1.9361111 0.2990961
# PSF   2.2748736 1.8301389 0.1226711
```

### 3.1. Adding New Error Metrics

As described in the earlier section, RMSE, MAPE, and MAE are the default error metrics assigned to ForecastTB. All these error metrics compare the forecasting methods in different perspectives [26]. Based on the information provided by each metric, users should use an appropriate one. All of these metrics are used internally within the prediction_errors() function.

Additionally, new error metrics can be added using an approach similar to that used for adding forecasting methods described in earlier sections. The following code demonstrates how the percentage change in variance (PCV, [47]) is added to the existing object as an additional error metric:

$$PCV = \frac{\lfloor var(Predicted) - var(Observed) \rfloor}{var(Observed)} \qquad (4)$$

where $var(Predicted)$ and $var(Observed)$ are variance of predicted and obvserved values.

The user-defined error function should be provided in an R script with two mandatory input arguments, that is, the same length vectors of observed and forecasted values. The defined function should return a summary of the errors or differences. The function and its name must be supplied to the ePara and ePara_name argument of the prediction_errors() function, respectively.

```r
# error metric to include with 'prediction_errors()' function
pcv <- function(obs, pred){
    d <- (var(obs) - var(pred)) * 100/ var(obs)
    d <- abs(as.numeric(d))
return(d)
}
```

The following code chunk shows how the new error function is attached to the prediction_errors object and corresponding update are observed with the plot() function, as shown in Figure 3.

```r
a1 <- prediction_errors(data = nottem, nval = 24,
Method = c("test1(data, nval)", "test2(data, nval)"),
MethodName = c("LPSF","PSF"),
ePara = "pcv(obs, pred)", ePara_name = 'PCV',
append_ = 1)
a1@output$Error_Parameters
```

| ## | RMSE | MAE | MAPE | PCV | exec_time |
|---|---|---|---|---|---|
| ## ARIMA | 2.8191449 | 2.3527391 | 4.9283784 | 17.3881625 | 0.1586161 |
| ## LPSF | 2.2180209 | 1.7208333 | 3.6520057 | 28.7072293 | 0.3137138 |
| ## PSF | 2.404563 | 1.781872 | 3.726139 | 23.872712 | 0.109726 |

```r
b1 <- plot(a1)
```

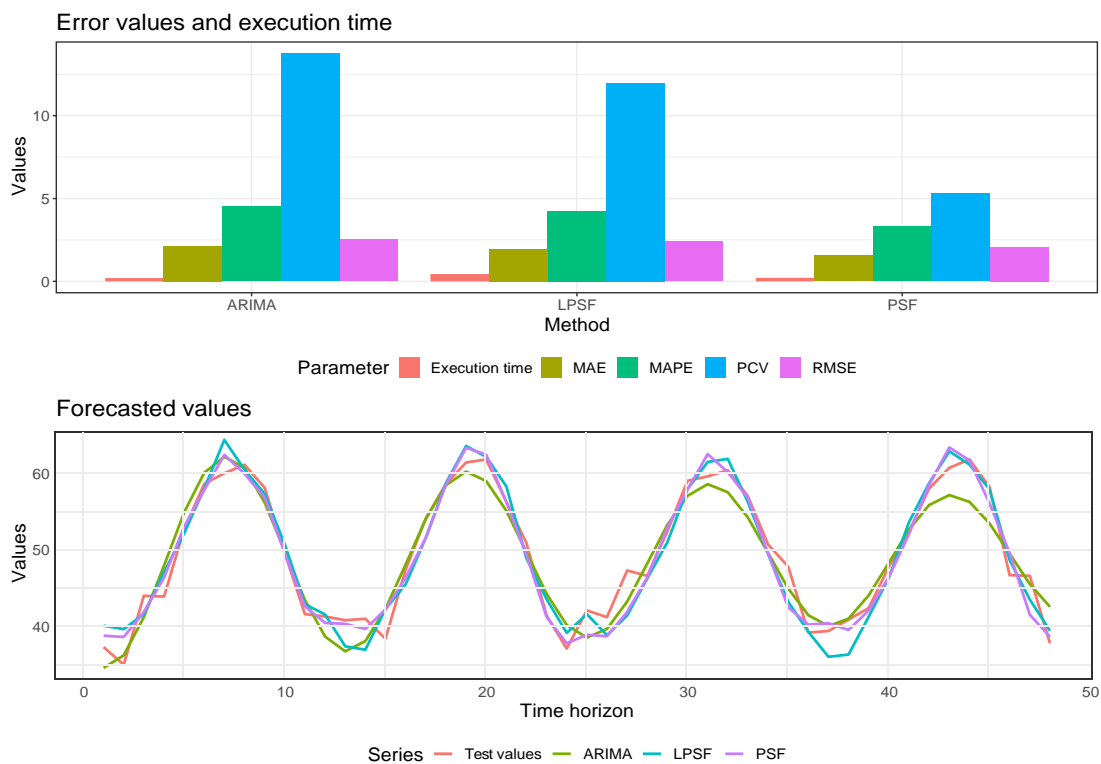

**Figure 3.** A plot showing the new error function in prediction_errors object.

### 3.2. A Polar Plot

The ForecastTB package provides a polar plot of showing forecasted values, especially if the time series dataset shows seasonality as shown in Figure 4. This plot shows the nature of forecasted values and how these values are behaving on an increasing number of time horizons. The numbers on the circle shows the forecasted time steps. Several forecasting methods are suitable for short-range forecasts but fail to maintain the accuracy for longer horizons. This proposed plot is capable of reflecting this feature in a more intuitive way. In Figure 4, the yellow-colored line for the ARIMA method is moving away from the test values as the time horizon proceeds.

```
plot_circle(a1)
```

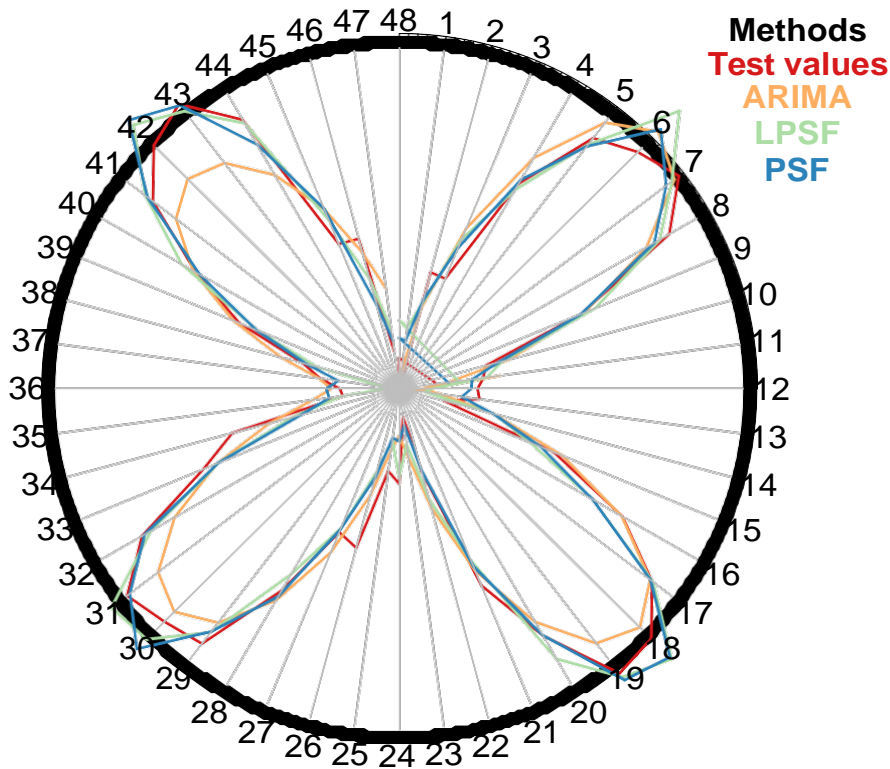

**Figure 4.** A polar plot to represent forecasted values.

### 3.3. Monte-Carlo Strategy

The code chunk below shows Monte-Carlo simulation on nottem time series with monte_carlo() function. This function is supplied with above mentioned prediction_errors object a1. It is also supplied with size of each data patch in each iteration as 144 and the number of iterations as 10. The flag parameters fval and figs are set to 0, in order to avoid display of forecasted values and comparison plots for all iterations in the Monte-Carlo simulations. The effect of these flags is further discussed in the vignette published in ForecastTB package.

```
monte_carlo(object = a1, size = 144, iteration = 10, fval = 0, figs = 0)
```

```
##          ARIMA      LPSF      PSF
## 19    3.597220  5.415615  5.228346
## 18    4.547199  5.411696  5.257050
## 67    2.080661  5.062059  4.747707
```

```
## 72   2.894787 5.392858 5.197034
## 79   2.641607 5.885326 5.003352
## 45   2.103611 4.739618 4.648975
## 58   2.522385 5.253795 5.140588
## 82   2.152816 4.889250 5.129120
## 33   3.193709 4.854308 4.949747
## 11   3.960024 5.834563 5.746631
## Mean 2.969402 5.273909 5.104855
```

This output shows the performance of ARIMA, LPSF, and PSF methods on nottem time series for 10 different time patches. The first column in the output shows the numeric number as an index in time series from which the data patch is selected for the analysis. Besides, the last row shows the averaged error values for all methods and it ensures more accurate and unbiased performance of the forecasting methods.

Furthermore, the following code chunk shows an object to compare the performance of forecasting methods on the sunspots time series and further a more rigorous comparison is done with Monte-Carlo simulations.

```r
x1 <- prediction_errors(data = sunspots, nval = 24,
Method = c("test1(data, nval)", "test2(data, nval)"),
MethodName = c("LPSF","PSF"),
append_ = 1)
x1@output$Error_Parameters
```

```
##           RMSE       MAE      MAPE exec_time
## ARIMA 67.141209 59.680909 95.074983  0.755336
## LPSF  45.251790 38.436458 37.283434  1.459494
## PSF   34.834718 28.803750 28.943332  0.496691
```

The flag parameters fval and figs in following monte_carlo() function are not provided, hence these flags takes the value 0 as default.

```r
monte_carlo(object = x1, size = 600, iteration = 10)
```

```
##           ARIMA     LPSF      PSF
## 2012  34.416174 24.383396 46.38433
## 252   20.804560 15.604257 15.56385
## 1524  14.008383 42.509944 22.56851
## 1081   5.928826 57.622837 23.16030
## 977   27.636099 35.118654 28.71433
## 734   16.529586 19.400491 21.04473
## 826   47.850499 26.893230 32.96957
## 1467  20.846583 17.438523 12.10345
## 484   29.408694 41.441812 31.73321
## 668    6.505468  7.243862 15.30992
## Mean  22.393487 28.765701 24.95522
```

It is interesting to know that, for the full-length sunspots time series, the PSF method was observed to be the best among the selected methods with the prediction_error() function, where last 24 values of the dataset were forecasted. Whereas, for the Monte-Carlo simulation, with rigorous iterations, ARIMA outperformed with a significant error difference as compared to PSF and LPSF methods. In this way, the monte_carlo() function can be used in comparing forecasting methods by avoiding the biased results that can be obtained by chance.

## 4. Case Study Related to the Energy Application

In the current research, it is important to validate the proposed R package ForecastTB based on real-world applications. In this manner, a reliable and feasible application can demonstrate the capacity of the developed R package forecasting code. Wind speed and solar radiation data of two meteorological stations (Robinson and Watford City) located in the North Dakota, USA, were used to test the predictability performance of the LPSF: modified pattern sequence-based forecast and PSF—pattern sequence-based forecast models. Three time scales were examined for the forecasting of 12 steps ahead of each scale including daily, weekly, and monthly. The modeling result was tested using several statistical performance metrics (RMSE, MAE, and MAPE), time convergence, and graphical presentation. The data span covers 20 years of historical data over the period (1 January 2000 to 31 December 2019). Three models were developed for modeling verification including support vector machine (SVM), decomposed pattern sequence-based forecast (DPSF), and hybridized ensemble empirical mode decomposition with Autoregressive integrated moving average (EEMD-ARIMA).

### 4.1. Statistical Performance Metrics Results

Following several established studies from the literature on wind speed and solar radiation prediction, RMSE, MAE, and MAPE are the commonly used statistical metrics for forecasting evaluation [48]. Those metrics can give an informative evaluation of the prediction capacity. For the Robinson station, the forecasting results attained the minimal values of the absolute error measures for the applied models—LPSF (RMSE = 7.57 and 71.6, MAE = 5.34 and 61.52 and MAPE = 35.84 and 28.22) and PSF (RMSE = 8.47 and 74.74, MAE = 7.0 and 63.65 and MAPE = 57.69 and 26.93) for the daily scale wind speed and solar radiation; LPSF (RMSE = 3.75 and 66.17, MAE = 3.2 and 47.36 and MAPE = 32.82 and 23.72) and PSF (RMSE = 3.3 and 59.33, MAE = 3.57 and 49.72 and MAPE = 24.5 and 21.88) for the weekly scale wind speed and solar radiation; LPSF (RMSE = 2.09 and 99.8, MAE = 1.58 and 82.74 and MAPE = 19.15 and 33.79) and PSF (RMSE = 1.6 and 39.32, MAE = 1.19 and 32.51 and MAPE = 13.77 and 10.84) for the monthly scale wind speed and solar radiation, respectively (Table 1). In comparison with the SVM, DPSF and EEMD-ARIMA, the LPSF and PSF models revealed an acceptable prediction enhancement. For instance, wind speed prediction was improved using the LPSF model by (86.04%, 42.77%, and 19.22%) based on the RMSE metric against the SVM, DPSF, and EEMD-ARIMA models, respectively. Whereas, solar radiation prediction accuracy was enhanced by (74.41%, 43.83% and 16.08%) using the LPSF model and RMSE metric over the SVM, DPSF and EEMD-ARIMA models, respectively.

**Table 1.** The statistical performance of the forecasting wind speed and solar radiation at Robinson station for the daily, weekly and monthly time scale for the applied (LSPF and PSF) and the comparable predictive models (i.e., SVM, DPSF and EEMD-ARIMA).

| Methods | Wind Speed | | | | Solar Radiation | | | |
|---|---|---|---|---|---|---|---|---|
| | **Daily Scale** | | | | | | | |
| | RMSE | MAE | MAPE | Execution Time | RMSE | MAE | MAPE | Execution Time |
| LPSF | 7.57 | 5.34 | 35.84 | 0.55 | 71.60 | 61.52 | 28.22 | 0.49 |
| PSF | 8.47 | 7.00 | 57.69 | 0.42 | 74.74 | 63.65 | 26.93 | 1.06 |
| SVM | 54.28 | 49.02 | 472.57 | 0.43 | 279.86 | 237.87 | 93.45 | 0.30 |
| DPSF | 13.24 | 11.07 | 73.07 | 1.12 | 127.47 | 115.48 | 42.43 | 1.63 |
| EEMD-ARIMA | 9.38 | 7.99 | 74.56 | 0.25 | 85.32 | 64.96 | 33.35 | 0.23 |

**Table 1.** *Cont.*

| Methods | Wind Speed | | | | Solar Radiation | | | |
|---|---|---|---|---|---|---|---|---|
| | **Weekly Scale** | | | | | | | |
| | **RMSE** | **MAE** | **MAPE** | **Execution Time** | **RMSE** | **MAE** | **MAPE** | **Execution Time** |
| LPSF | 3.75 | 3.20 | 32.82 | 0.19 | 66.17 | 47.36 | 23.72 | 0.30 |
| PSF | 3.30 | 2.57 | 24.50 | 0.50 | 59.33 | 49.72 | 21.88 | 0.78 |
| SVM | 13.51 | 10.73 | 117.89 | 0.30 | 548.47 | 438.63 | 250.93 | 0.24 |
| DPSF | 3.40 | 2.74 | 24.28 | 1.64 | 260.68 | 238.28 | 128.00 | 1.77 |
| EEMD-ARIMA | 2.73 | 2.33 | 24.60 | 0.44 | 147.88 | 124.65 | 67.65 | 0.39 |
| | **Monthly Scale** | | | | | | | |
| | **RMSE** | **MAE** | **MAPE** | **Execution Time** | **RMSE** | **MAE** | **MAPE** | **Execution Time** |
| LPSF | 2.09 | 1.58 | 19.15 | 0.47 | 99.80 | 82.74 | 33.79 | 0.46 |
| PSF | 1.60 | 1.19 | 13.77 | 0.35 | 39.32 | 32.51 | 10.84 | 0.32 |
| SVM | 8.66 | 7.11 | 77.86 | 0.18 | 3881.20 | 3384.89 | 1632.05 | 0.19 |
| DPSF | 1.73 | 1.49 | 16.65 | 0.95 | 100.60 | 88.72 | 32.25 | 1.02 |
| EEMD-ARIMA | 1.93 | 1.63 | 19.55 | 0.19 | 50.00 | 36.94 | 13.90 | 3.62 |

For Watford City station, the best forecasting results achieved for the proposed LPSF and PSF models with minimum values of (RMSE, MAE, and MAPE). Statistically, the performance metrics accomplished were; LPSF (RMSE = 6.67 and 162.74, MAE = 5.48 and 158.21 and MAPE = 39.85 and 156.09) and PSF (RMSE = 6.98 and 131.67, MAE = 5.73 and 123.64 and MAPE = 41.45 and 125.34) for the daily scale wind speed and solar radiation; LPSF (RMSE = 2.95 and 70.21, MAE = 2.04 and 55.65 and MAPE = 23.68 and 26.74) and PSF (RMSE = 2.18 and 80.83, MAE = 1.79 and 68.79 and MAPE = 21.33 and 32.76) for the weekly scale wind speed and solar radiation; LPSF (RMSE = 1.98 and 115.18, MAE = 1.7 and 90.65 and MAPE = 20.16 and 42.48) and PSF (RMSE = 1.43 and 32.09, MAE = 1.17 and 25.23 and MAPE = 13.84 and 8.77) for the monthly scale wind speed and solar radiation, respectively (Table 2). The statistical results of the Watford City station demonstrated more or less similar predictability performance of wind speed and solar radiation in comparison with the benchmark models (i.e., SVM, DPSF and EEMD-ARIMA).

**Table 2.** The statistical performance of the forecasting wind speed and solar radiation at Watford City station for the daily, weekly and monthly time scale for the applied (LSPF and PSF) and the comparable predictive models (i.e., SVM, DPSF and EEMD-ARIMA).

| Methods | Wind Speed | | | | Solar Radiation | | | |
|---|---|---|---|---|---|---|---|---|
| | **Daily Scale** | | | | | | | |
| | **RMSE** | **MAE** | **MAPE** | **Execution Time** | **RMSE** | **MAE** | **MAPE** | **Execution Time** |
| LPSF | 6.67 | 5.48 | 39.85 | 0.48 | 162.74 | 158.21 | 156.09 | 0.23 |
| PSF | 6.98 | 5.73 | 41.45 | 0.67 | 131.67 | 123.64 | 125.34 | 1.69 |
| SVM | 39.51 | 36.36 | 377.76 | 0.26 | 235.05 | 198.34 | 192.34 | 0.26 |
| DPSF | 10.48 | 9.02 | 64.90 | 1.77 | 44.29 | 36.30 | 30.20 | 1.18 |
| EEMD-ARIMA | 9.20 | 8.20 | 64.60 | 0.27 | 33.86 | 29.13 | 29.67 | 0.54 |

**Table 2.** *Cont.*

| Methods | Wind Speed | | | | Solar Radiation | | | |
|---|---|---|---|---|---|---|---|---|
| | **Weekly Scale** | | | | | | | |
| | **RMSE** | **MAE** | **MAPE** | **Execution Time** | **RMSE** | **MAE** | **MAPE** | **Execution Time** |
| LPSF | 2.95 | 2.04 | 23.68 | 0.79 | 70.21 | 55.65 | 26.74 | 0.22 |
| PSF | 2.18 | 1.79 | 21.33 | 0.44 | 80.83 | 68.79 | 32.76 | 0.96 |
| SVM | 11.05 | 8.60 | 94.95 | 0.26 | 117.46 | 89.32 | 53.93 | 0.53 |
| DPSF | 2.28 | 1.61 | 17.26 | 2.16 | 134.15 | 124.54 | 61.43 | 1.21 |
| EEMD-ARIMA | 1.87 | 1.55 | 17.56 | 0.38 | 147.10 | 126.24 | 71.94 | 0.23 |
| | **Monthly Scale** | | | | | | | |
| | **RMSE** | **MAE** | **MAPE** | **Execution Time** | **RMSE** | **MAE** | **MAPE** | **Execution Time** |
| LPSF | 1.98 | 1.70 | 20.16 | 1.30 | 115.18 | 90.65 | 42.48 | 1.05 |
| PSF | 1.43 | 1.17 | 13.84 | 0.71 | 32.09 | 25.23 | 8.77 | 0.49 |
| SVM | 3.34 | 2.74 | 31.63 | 0.37 | 3739.78 | 3231.79 | 1719.36 | 0.52 |
| DPSF | 1.90 | 1.67 | 19.46 | 1.79 | 102.93 | 92,20 | 35.60 | 1.76 |
| EEMD-ARIMA | 1.30 | 1.02 | 11.29 | 0.73 | 65.96 | 58.27 | 26.29 | 0.39 |

The convergence execution time is one of the essential modeling components in the field of soft computing and modeling aspects [49]. With respect to the convergence execution time of the developed models, the LPSF and PSF revealed a very faster time learning process with an acceptable execution process.

### 4.2. The Graphical Presentation of the Developed Forecasting Models

Figure 5a–c reported the performance metrics and time execution results of the wind speed forecasting daily, weekly, and monthly in the form of a bar chart. For all investigated time scales, the LPSF and PSF models revealed a slightly similar performance to the DPSF and EEMD-ARIMA. However, the SVM model showed very poor predictability performance and this can be explained due to the essential tuning procedure required for the SVM model for the internal parameters, that were not adopted or focused on in this research [50,51]. Figure 6a–c indicated the performance metrics and time execution results of solar radiation forecasting for the daily, weekly, and monthly scales at the Robinson station. It is clearly seen that the monthly scale solar radiation for LPSF, PSF, DPSF, and EEMD-ARIMA models were similar. However, the SVM model still reported the lowest capacity. Figure 7a–c displayed the forecasting capacity of the proposed and comparable forecasting models for wind speed modeling at Watford City station. For the daily and weekly wind speed scales, the LPSF and PSF models exhibited a distinguished performance against the results of SVM models. Yet, the monthly scale results of the LPSF, and PSF models were slightly poorer and similar to the ones attained by DPSF and EEMD-ARIMA. Figure 8a–c reported the performance of the solar radiation forecasting Watford City station and for all molded time series scales. For the daily scale, Figure 8a shows that the performance of the DPSF and EEMD-ARIMA models is superior to that of the LPSF models and PSF models. This might be due to the high stochasticity of the solar radiation process on a daily scale and thus the proposed models could not capture the actual trend. On the weekly base scale, the DPSF and EEMD-ARIMA models indicated the worst forecasting potential, even worse than the SVM model. Whereas, similar performance attained using LPSF, PSF, DPSF, and EEMD-ARIMA models for the monthly scale of solar radiation other hand, for the monthly.

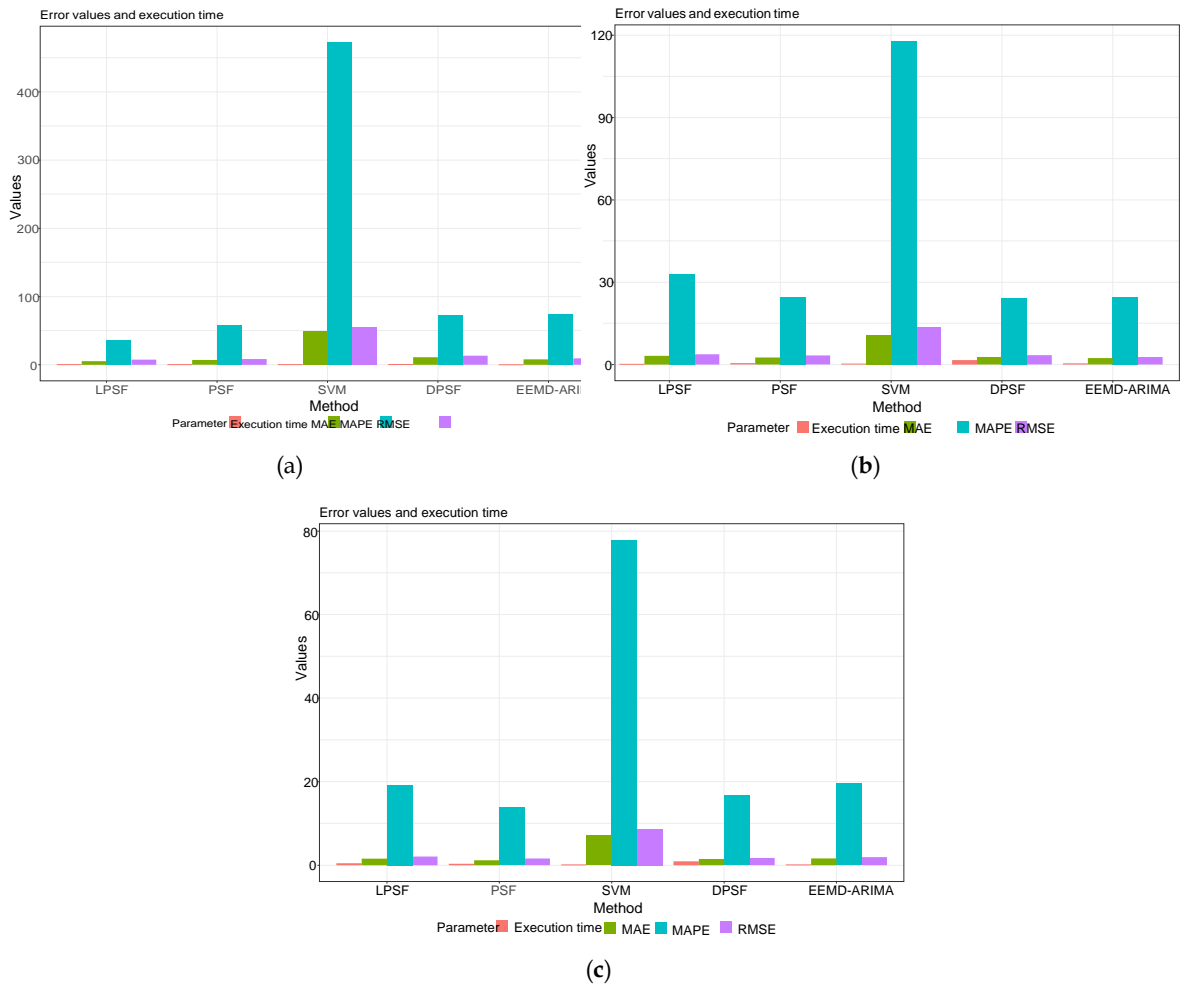

(a)

(b)

(c)

**Figure 5.** The bar chart of the performance metrics and time execution results of wind speed forecasting for the daily (**a**), weekly (**b**) and monthly (**c**) at Robinson station.

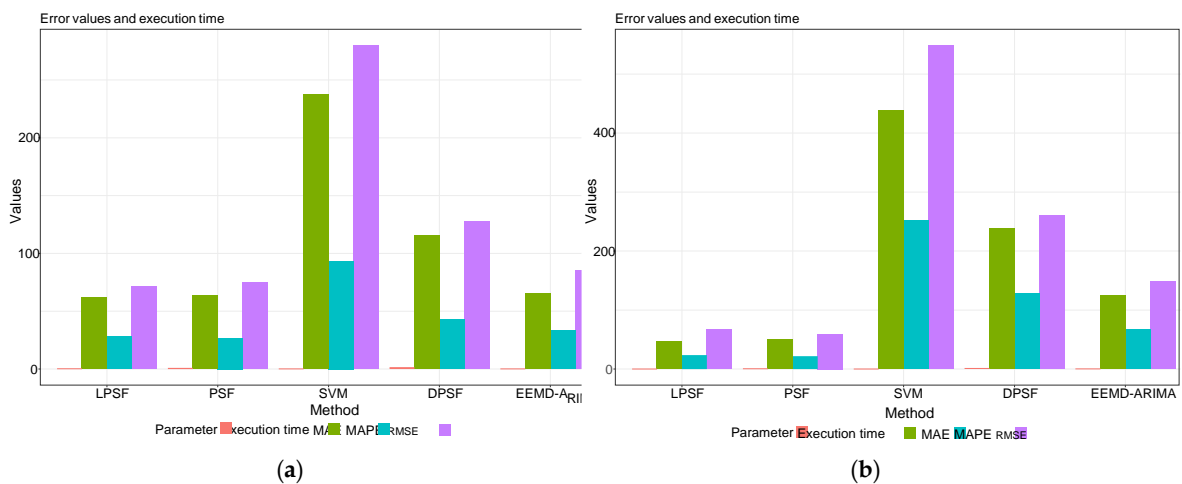

(a)

(b)

**Figure 6.** *Cont.*

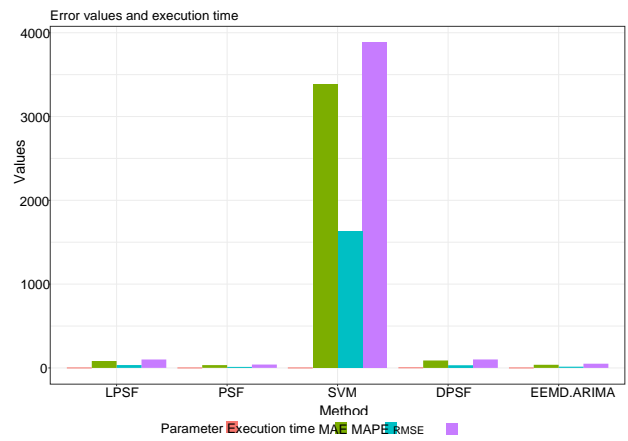

(**c**)

**Figure 6.** The bar chart of the performance metrics and time execution results of solar radiation forecasting for the daily (**a**), weekly (**b**) and monthly (**c**) at Robinson station.

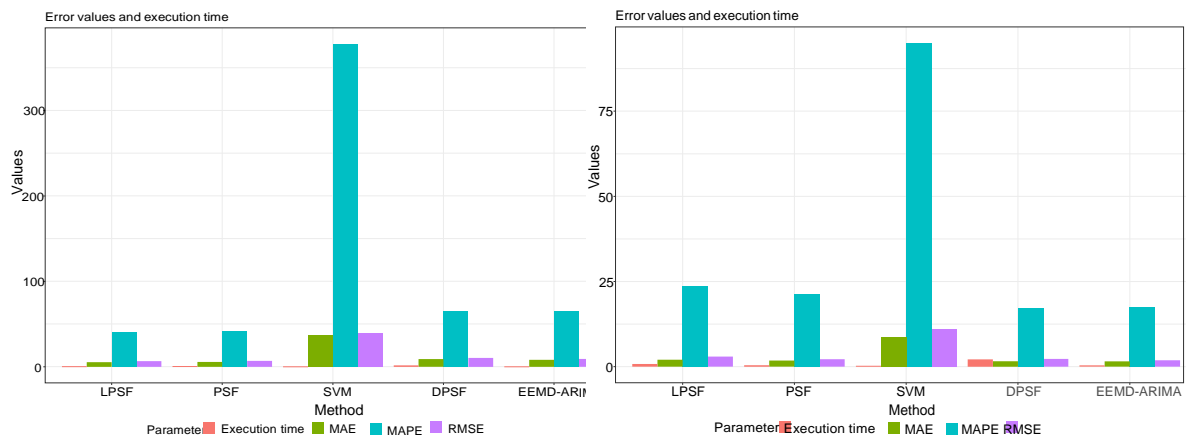

(**a**)                                                                (**b**)

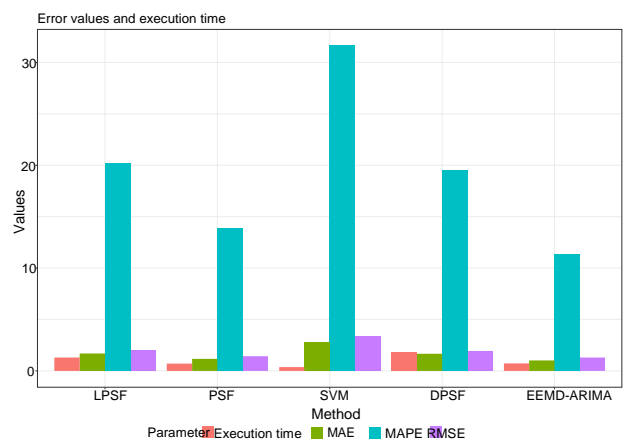

(**c**)

**Figure 7.** The bar chart of the performance metrics and time execution results of wind speed forecasting for the daily (**a**), weekly (**b**) and monthly (**c**) at Watford City station.

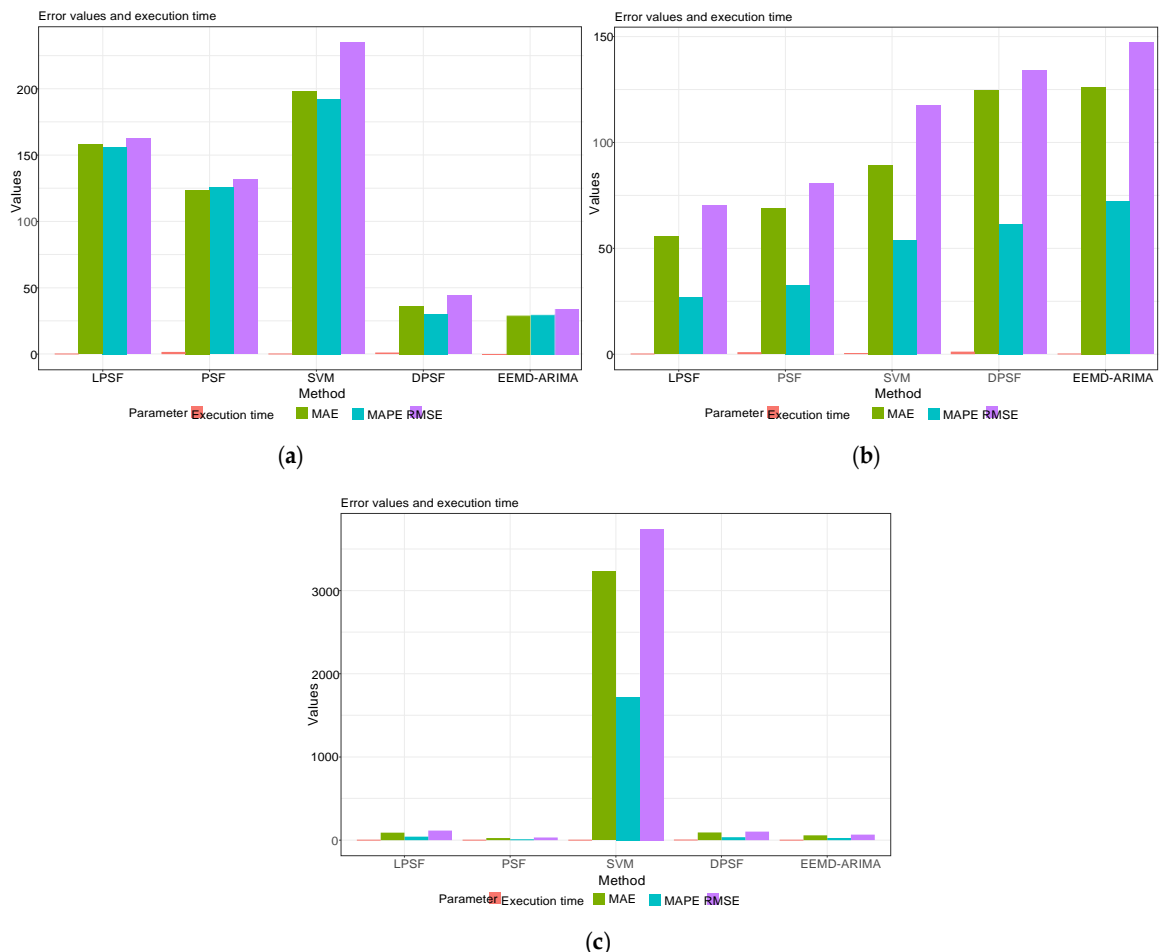

**Figure 8.** The bar chart of the performance metrics and time execution results of wind speed forecasting for the daily (**a**), weekly (**b**) and monthly (**c**) at Watford City station.

## 5. Discussion

This article proposed and demonstrated the ForecastTB package as a test-bench for comparing the time series forecasting methods as a crucial step towards more formal time series analysis. This demonstration is further described with some case studies to show how characteristics of the temporal correlation structure can influence the forecast accuracy. The ForecastTB package greatly assists in comparing different forecasting methods and considering the characteristic of the time series dataset.

Also, this paper explains how new forecasting methods and error metrics can be introduced in the comparative test-bench. As such, the package allows users to include additional forecasting methods for comparison, which can be extremely useful given the capability of R to interface with other programming languages (e.g., matlabr for MatLab [52]; Rcpp for compiled languages [53] and many other). Finally, a simple plug-and-play module based architecture of the ForecastTB to append or remove several forecasting methods and error metrics makes it a robust and handy tool to evaluate forecasting analysis.

It is worth mentioning that forecasting different time scale wind speed and solar radiation is highly essential for multiple energy sources harvesting [54–56]. Providing an accurate and reliable forecasting intelligent computer aid models for wind speed and solar radiation can contribute to the base knowledge of friendly energy sources and sustainable green cities. To the best knowledge of the current research, the feasibility of the LPSF and PSF models is evidenced to be capable to module related energy datasets.

**Author Contributions:** All the authors contributed equally to this work. Conceptualization, N.D.B. and G.B.A.; methodology, N.D.B. and G.B.A.; software, N.D.B.; validation, N.D.B. and Z.M.Y.; formal analysis, N.D.B., Z.M.Y. and G.B.A.; investigation, N.D.B.; resources, Z.M.Y. and G.B.A.; data curation, N.D.B. and Z.M.Y.; writing—original draft preparation, N.D.B.; writing—review and editing, G.B.A.; visualization, N.D.B.; supervision, G.B.A.; project administration, G.B.A.; funding acquisition, G.B.A. All authors have read and agreed to the published version of the manuscript.

**Funding:** This study was funded by Apple Inc. as part of the APPLAUSE bio-energy collaboration with Aarhus University.

**Acknowledgments:** Authors acknowledge the R community and The Comprehensive R Archive Network (CRAN) maintainers for providing feedback over the package.

**Conflicts of Interest:** The authors declare no conflict of interest.

## Abbreviations

The following abbreviations are used in this manuscript:

| | |
|---|---|
| ARIMA | Auto Regressive Integrated Moving Average |
| CRAN | The Comprehensive R Archive Network |
| ETS | Error, Trend, Seasonal |
| LPSF | Modified Pattern Sequence based Forecast |
| MATLAB | Matrix Laboratory |
| MAE | Mean Absolute Error |
| MAPE | Mean Absolute Percentage Error |
| PCV | Percent Change in Variance |
| PSF | Pattern Sequence based Forecast |
| RMSE | Root Mean Square Error |